\documentclass[10pt,conference]{IEEEtran}

\usepackage{graphicx}
\usepackage{subfigure}
\usepackage{cite}
\usepackage{amssymb, amsmath}
\usepackage[flushleft]{paralist}

\IEEEoverridecommandlockouts

\begin{document}

\bstctlcite{TurboRef:BSTcontrol}

\title{Can Punctured Rate-1/2 Turbo Codes Achieve a Lower Error Floor than their Rate-1/3 Parent Codes?}

\author{\authorblockN{Ioannis Chatzigeorgiou, Miguel R. D. Rodrigues, Ian J. Wassell}
\authorblockA{Digital~Technology~Group,~Computer~Laboratory\\
University~of~Cambridge,~United~Kingdom\\
Email:~\{ic231,~mrdr3,~ijw24\}@cam.ac.uk} \and
\authorblockN{Rolando Carrasco}
\authorblockA{Department~of~EE\&C~Engineering\\
University~of~Newcastle,~United~Kingdom\\
Email:~r.carrasco@ncl.ac.uk}
\thanks{This work was supported by EPSRC
under Grant GR/S46437/01.}}

\maketitle
\begin{abstract}
In this paper we concentrate on rate-1/3 systematic parallel
concatenated convolutional codes and their rate-1/2 punctured child
codes. Assuming maximum-likelihood decoding over an additive white
Gaussian channel, we demonstrate that a rate-1/2 non-systematic
child code can exhibit a lower error floor than that of its rate-1/3
parent code, if a particular condition is met. However, assuming
iterative decoding, convergence of the non-systematic code towards
low bit-error rates is problematic. To alleviate this problem, we
propose rate-1/2 partially-systematic codes that can still achieve a
lower error floor than that of their rate-1/3 parent codes. Results
obtained from extrinsic information transfer charts and simulations
support our conclusion.
\end{abstract}

\IEEEpeerreviewmaketitle

\section{Introduction}
\label{Intro}

A punctured convolutional code is obtained by the periodic
elimination of symbols from the output of a low-rate parent
convolutional code. Extensive analyses on the structure and
performance of punctured convolutional codes has shown that their
performance is always inferior than the performance of their
low-rate parent codes (e.g. see \cite{Hagenauer88,Haccoun89}).

The performance of punctured parallel concatenated convolutional
codes (PCCCs), also known as punctured turbo codes, has also been
investigated. Design considerations have been derived by analytical
\cite{Acikel99,Babich02,Kousa02} as well as simulation-based
approaches \cite{FanMo99,Land00,Blazek02}, while upper bounds to the
bit error probability (BEP) were evaluated in
\cite{Kousa02,Chatzigeorgiou06b}. The most recent papers
\cite{Land00,Blazek02,Chatzigeorgiou06b} demonstrate that puncturing
both systematic and parity outputs of a rate-1/3 turbo code results
in better high-rate turbo codes, in terms of BEP performance, than
puncturing only the parity outputs of the original turbo code.

The aim of this paper is to explore whether rate-1/2 punctured turbo
codes can eventually achieve better performance than their parent
rate-1/3 systematic turbo codes on additive white Gaussian (AWGN)
channels. Assuming maximum-likelihood (ML) decoding, we demonstrate
that, contrary to punctured convolutional codes, punctured turbo
codes yielding lower error floors than that of their parent code can
be constructed. Nevertheless, we cannot be conclusive when
suboptimal iterative decoding is used. For this reason, we also
study the convergence behavior of iterative decoding and we
investigate whether the performance of the proposed rate-1/2
punctured turbo codes converges towards the theoretical error floor,
at low bit error probabilities.

\section{Performance Evaluation of\\Rate-1/3 Turbo Codes}
\label{Definitions}

Turbo codes, in the form of symmetric rate-1/3 PCCCs, consist of two
identical rate-1/2 recursive systematic convolutional encoders
separated by an interleaver of size $N$ \cite{Berrou96}. The
information bits are input to the first constituent convolutional
encoder, while an interleaved version of the information bits are
input to the second convolutional encoder. The output of the turbo
encoder consists of the systematic bits of the first encoder, which
are identical to the information bits, the parity check bits of the
first encoder and the parity check bits of the second encoder.

It was shown in \cite{Divsalar95} and \cite{Benedetto96a} that the
performance of a PCCC can be obtained from the input-redundancy
weight enumerating functions (IRWEFs) of the terminated constituent
recursive convolutional codes. The IRWEF for the case of a
convolutional code $\mathcal{C}$ assumes the form
\begin{equation}
\label{C_IRWEF}
A^{\mathcal{C}}(W,Z)=\:\sum\limits_{w}\sum\limits_{j}
A^{\mathcal{C}}_{w,j}W^{w}Z^{j},
\end{equation}where $A^{\mathcal{C}}_{w,j}$ denotes the number of codeword
sequences having parity check weight $j$, which were generated by an
input sequence of weight $w$. The overall output weight of the
codeword sequence, for the case of a systematic code, is $w+j$.

The conditional weight enumerating function (CWEF),
$A^{\mathcal{C}}(w,Z)$, provides all codeword sequences generated by
an input sequence of weight $w$. Consequently, the relationship
between the CWEF and the IRWEF is
\begin{equation}
\label{C_CWEF} A^{\mathcal{C}}(W,Z)=\:\sum\limits_{w}
A^{\mathcal{C}}(w,Z)W^{w}.
\end{equation}

A relationship between the CWEF of a PCCC, $\mathcal{P}$, and the
CWEF of $\mathcal{C}$, which is one of the two identical constituent
codes, can be easily derived only if we assume the use of a uniform
interleaver, an abstract probabilistic concept introduced in
\cite{Benedetto96a}. In particular, if $N$ is the size of the
uniform interleaver and $A^{\mathcal{C}}(w,Z)$ is the CWEF of the
constituent code, the CWEF of the PCCC, $A^{\mathcal{P}}(w,Z)$, is
equal to
\begin{equation}
\label{PCCC_CWEF} A^{\mathcal{P}}(w,Z) =
\frac{\left[A^{\mathcal{C}}(w,Z)\right]^{2}} {\displaystyle
\binom{N}{w}}.
\end{equation}The IRWEF of $\mathcal{P}$, $A^{\mathcal{P}}(W,Z)$, can be then computed from the CWEF, $A^{\mathcal{P}}(w,Z)$,
in a manner identical to (\ref{C_CWEF}).

The input-output weight enumerating function (IOWEF) provides the
number of codeword sequences generated by an input sequence of
weight $w$, whose overall output weight is $d$, in contrast with the
IRWEF, which only considers the output parity check weight $j$. If
$\mathcal{P}$ is a systematic PCCC, the corresponding IOWEF assumes
the form
\begin{equation}
\label{PCCC_IOWEF}
B^{\mathcal{P}}(W,D)=\:\sum\limits_{w}\sum\limits_{d}B^{\mathcal{P}}_{w,d}W^{w}D^{d},
\end{equation} where the coefficients $B^{\mathcal{P}}_{w,d}$ can be derived
from the coefficients $A^{\mathcal{P}}_{w,z}$ of the IRWEF, based on
the expressions
\begin{equation}
\label{PCCC_IOWEF_COEFF}
B^{\mathcal{P}}_{w,d}=A^{\mathcal{P}}_{w,j},\quad \text{and}\quad
d=w+j.
\end{equation}

The IOWEF coefficients $B^{\mathcal{P}}_{w,d}$ can be used to
determine a tight upper bound on the BEP for ML soft decoding for
the case of an AWGN channel, as follows
\begin{equation}
\label{PB_old} P_{B} \leq \frac{1}{N}
\sum\limits_{d}\sum\limits_{w}wB^{\mathcal{P}}_{w,d}
Q\left(\sqrt{\frac{2R_{\mathcal{P}}E_{b}}{N_{0}}\cdot d} \right),
\end{equation}where $R_{\mathcal{P}}$ is the rate of the turbo code, which
in our case is equal to $1/3$. The upper bound can be rewritten as
\begin{equation}
\label{PB} P_{B} \leq \sum\limits_{w}P(w),
\end{equation}where $P(w)$ is the contribution to the overall BEP of all error events
having information weight $w$, and is defined as
\begin{equation}
\label{Pw} P(w)=\:\sum\limits_{d}\frac{w}{N}B^{\mathcal{P}}_{w,d}
Q\left(\sqrt{\frac{2R_{\mathcal{P}}E_{b}}{N_{0}}\cdot d} \right).
\end{equation}

Benedetto \textit{et al.} showed in \cite{Benedetto96a} that the
upper bound on the BEP of a PCCC using a uniform interleaver of size
$N$ coincides with the average of the upper bounds obtainable from
the whole class of deterministic interleavers of size $N$. For small
values of $N$, the upper bound can be very loose compared with the
actual performance of turbo codes using specific deterministic
interleavers. However, for $N\!\geq\!1000$, it has been observed
that randomly generated interleavers generally perform better than
deterministic interleaver designs \cite{Hall98}. Consequently, the
upper bound provides a good indication of the actual error rate
performance of a PCCC, when long interleavers are considered.

\section{Performance Evaluation of\\Rate-1/2 Punctured Non-Systematic PCCCs}
\label{Perf_Rate12}

Rates higher than 1/3 can be achieved by puncturing the output of a
rate-1/3 turbo encoder. Punctured codes are classified as systematic
(S), partially systematic (PS) or non-systematic (NS) depending on
whether all, some or none of their systematic bits are transmitted
\cite{Land00}. In this section we concentrate specifically on
rate-1/2 NS-PCCCs, because their weight enumerating functions can be
easily related to the weight enumerating functions of their parent
codes, as it will now be demonstrated.

A symmetric rate-1/2 NS-PCCC, $\mathcal{P^{\prime}}$, can be
obtained by puncturing the systematic output of a rate-1/3 PCCC,
$\mathcal{P}$, which consists of two identical rate-1/2 recursive
systematic convolutional codes. However, $\mathcal{P^{\prime}}$ can
also be seen as a PCCC constructed using two identical rate-1
non-systematic convolutional codes, each one of which has been
obtained by puncturing the systematic bits of a rate-1/2 systematic
convolutional code, identical to the one used in $\mathcal{P}$. If
$\mathcal{C^{\prime}}$ is the punctured rate-1 non-systematic
convolutional code and $\mathcal{C}$ is the parent rate-1/2
systematic convolutional code, their IRWEFs,
$A^{\mathcal{C^{\prime}}}(W,Z)$ and $A^{\mathcal{C}}(W,Z)$
respectively, are identical, i.e.,
\begin{equation}
\label{NS-C_IRWEF}
A^{\mathcal{C^{\prime}}}(W,Z)=A^{\mathcal{C}}(W,Z),
\end{equation}since, by definition, the IRWEF does not provide information about the
weight of the systematic bits of a codeword. Thus, puncturing of the
systematic bits of $\mathcal{C}$ will not cause any change in its
IRWEF.

Either by applying the same reasoning or by considering
(\ref{C_CWEF}) and (\ref{PCCC_CWEF}), we find that the IRWEF of the
rate-1/2 NS-PCCC, $A^{\mathcal{P^{\prime}}}(W,Z)$, is identical to
the IRWEF of its parent code, $A^{\mathcal{P}}(W,Z)$, i.e.,
\begin{equation}
\label{NS-PCCC_IRWEF}
A^{\mathcal{P^{\prime}}}(W,Z)=A^{\mathcal{P}}(W,Z).
\end{equation}

Puncturing has an effect only when calculating the IOWEF of
$\mathcal{P^{\prime}}$, $B^{\mathcal{P^{\prime}}}(W,D)$. We use the
notation $d^{\prime}$ to denote the overall weight of a codeword
sequence after puncturing as opposed to $d$, which refers to the
overall weight of the same codeword sequence before puncturing.
Therefore the IOWEF of $\mathcal{P^{\prime}}$ can be expressed as
\begin{equation}
\label{NS-PCCC_IOWEF}
B^{\mathcal{P^{\prime}}}(W,D)=\:\sum\limits_{w}\sum\limits_{d^{\prime}}B^{\mathcal{P^{\prime}}}_{w,d^{\prime}}W^{w}D^{d^{\prime}}.
\end{equation}Since all systematic bits are punctured, the weight $w$ of the information bits does not
contribute to the overall weight $d^{\prime}$ of the punctured
codeword sequences, and hence it follows that
\begin{equation}
\label{NS-PCCC_IOWEF_COEFF}
B^{\mathcal{P^{\prime}}}_{w,d^{\prime}}=A^{\mathcal{P}}_{w,j},\quad
\text{and}\quad d^{\prime}=j.
\end{equation}From (\ref{PCCC_IOWEF_COEFF}) and (\ref{NS-PCCC_IOWEF_COEFF}) we find that the relationship between the IOWEF coefficients, $B^{\mathcal{P^{\prime}}}_{w,d^{\prime}}$ and $B^{\mathcal{P}}_{w,d}$, is as follows
\begin{equation}
\label{COEFF_Relationship}
B^{\mathcal{P^{\prime}}}_{w,d^{\prime}}=B^{\mathcal{P}}_{w,(d^{\prime}+w)}\quad
\text{or}\quad
B^{\mathcal{P^{\prime}}}_{w,(d-w)}=B^{\mathcal{P}}_{w,d}
\end{equation}since
\begin{equation}
\label{OUTPUT_WEIGHT}
d=d^{\prime}+w.
\end{equation}This is equivalent to saying that if all information sequences of weight $w$ are input to both $\mathcal{P}$ and $\mathcal{P^\prime}$, the overall weight of the generated codeword sequences follows the same distribution in both cases,
but is shifted by $w$ in the case of $\mathcal{P^\prime}$.


The upper bound on the BEP for ML soft decoding for the case of an
AWGN channel takes the form
\begin{equation}
\label{NS-PB} P_{B} \leq \sum\limits_{w}P^{\prime}(w),
\end{equation}where $P^{\prime}(w)$ is given by
\begin{equation}
\label{NS-Pw}
P^{\prime}(w)=\:\sum\limits_{d^{\prime}}\frac{w}{N}B^{\mathcal{P^{\prime}}}_{w,d^{\prime}}
Q\left(\sqrt{\frac{2R_{\mathcal{P^{\prime}}}E_{b}}{N_{0}}\cdot
d^{\prime}} \right)
\end{equation} and $R_\mathcal{P^{\prime}}=1/2$. Taking into account (\ref{COEFF_Relationship}) and (\ref{OUTPUT_WEIGHT}), we can rewrite (\ref{NS-Pw}) as a function of $d$ and $B^{\mathcal{P}}_{w,d}$, as follows
\begin{equation}
\label{new_NS-Pw}
P^{\prime}(w)=\:\sum\limits_{d}\frac{w}{N}B^{\mathcal{P}}_{w,d}
Q\left(\sqrt{\frac{2R_{\mathcal{P^{\prime}}}E_{b}}{N_{0}}\cdot
\left(d-w\right)} \right)
\end{equation}


\section{Performance Analysis}
\label{Conditions}

Before continuing to the derivation of a condition which needs to be
met so that a rate-1/2 NS-PCCC can achieve a better ML bound than
its parent rate-1/3 PCCC, we first enumerate a number of results,
derived and justified in \cite{Benedetto96b}:
\begin{enumerate}
  \item The minimum information weight $w_{\text{min}}$ for recursive convolutional encoders is $w_{\text{min}}=2$.
  \item For recursive constituent codes and long interleavers, the contribution to the overall BEP of all error events
with odd information weight is negligible. Furthermore, as the interleaver size increases, the contribution to the overall BEP of all error events with
information weight $w_{\text{min}}$ is dominant.
  \item The upper bound of a PCCC which uses recursive constituent codes, depends on its free effective distance $d_{\text{free.eff}}$,
which corresponds to the minimum overall output weight when the information weight is $w_{\text{min}}$.
\end{enumerate}
It is also straightforward to verify that, although puncturing of
the systematic bits of a rate-1/3 PCCC affects the upper bound on
the BEP, the previously highlighted trends still apply.

Based on (\ref{PB}) and (\ref{NS-PB}), the rate-1/2 NS-PCCC,
$\mathcal{P^{\prime}}$, will achieve a better bound than its parent,
if
\begin{equation}
\label{Find_Condition_1} \sum\limits_{w\geq2}P^{\prime}(w) <
\sum\limits_{w\geq2}P(w).
\end{equation}For large interleaver sizes, the dominant terms will be $P^{\prime}(2)$ and $P(2)$, thus (\ref{Find_Condition_1}) reduces
to $P^{\prime}(2)<P(2)$, or equivalently
\begin{equation}
\label{Find_Condition_2}
\sum\limits_{d\geq
d_{\text{free.eff}}}Q\left(f^{\prime}\left(d\right)\right)
<\sum\limits_{d\geq
d_{\text{free.eff}}}Q\left(f\left(d\right)\right),
\end{equation}where $f^{\prime}(d)$ and $f(d)$ are defined as
\begin{equation}
\label{Find_Condition_3}
\begin{split}
f^{\prime}(d)&=\sqrt{\frac{2R_{\mathcal{P^{\prime}}}E_{b}}{N_{0}}\cdot \left(d-2\right)},\\
f(d)&=\sqrt{\frac{2R_{\mathcal{P}}E_{b}}{N_{0}}\cdot d},
\end{split}
\end{equation}according to (\ref{Pw}) and
(\ref{new_NS-Pw}).


Function $Q(\xi)$ is a monotonically decreasing function of $\xi$,
where $\xi$ is a real number. Consequently, if $\xi_{1}$ and
$\xi_{2}$ are real numbers with $\xi_{1}>\xi_{2}$, it follows that
$Q(\xi_{1})<Q(\xi_{2})$, and vice versa, i.e,
\begin{equation}
\label{Q_Condition} Q(\xi_{1})<Q(\xi_{2})\Leftrightarrow
\xi_{1}>\xi_{2}.
\end{equation}Therefore,
inequality (\ref{Find_Condition_2}) is satisfied if
\begin{equation}
\label{Find_Condition_4} f^{\prime}(d)>f(d),\quad\text{for
every}\;d\geq d_{\text{free.eff}}.
\end{equation}Both $f^{\prime}(d)$
and $f(d)$ are monotonically increasing functions, therefore
(\ref{Find_Condition_4}) holds true if only
$f^{\prime}(d_{\text{free.eff}})>f(d_{\text{free.eff}})$, or
\begin{equation}
\label{Find_Condition_5} d_{\text{free.eff}} >
\frac{2R_{\mathcal{P^{\prime}}}}{R_{\mathcal{P^{\prime}}}-R_{\mathcal{P}}}\:.
\end{equation}After substituting $R_{\mathcal{P^{\prime}}}$ and $R_{\mathcal{P}}$
with $1/2$ and $1/3$ respectively, we find that a rate-1/2 NS-PCCC
can achieve a lower upper bound on the BEP, over an AWGN channel,
than its rate-1/3 parent PCCC only if the effective free distance of
the parent PCCC, $d_{\text{free.eff}}$, meets the condition
\begin{equation}
\label{Find_Condition_6}d_{\text{free.eff}}>6.
\end{equation}


\begin{figure}[t]
    \centering
    \setlength{\abovecaptionskip}{3pt}
    \setlength{\belowcaptionskip}{0pt}
    \begin{tabular}{l}
        \begin{minipage}[b]{0.95\linewidth}
            \centering
            \includegraphics[width=1\linewidth]{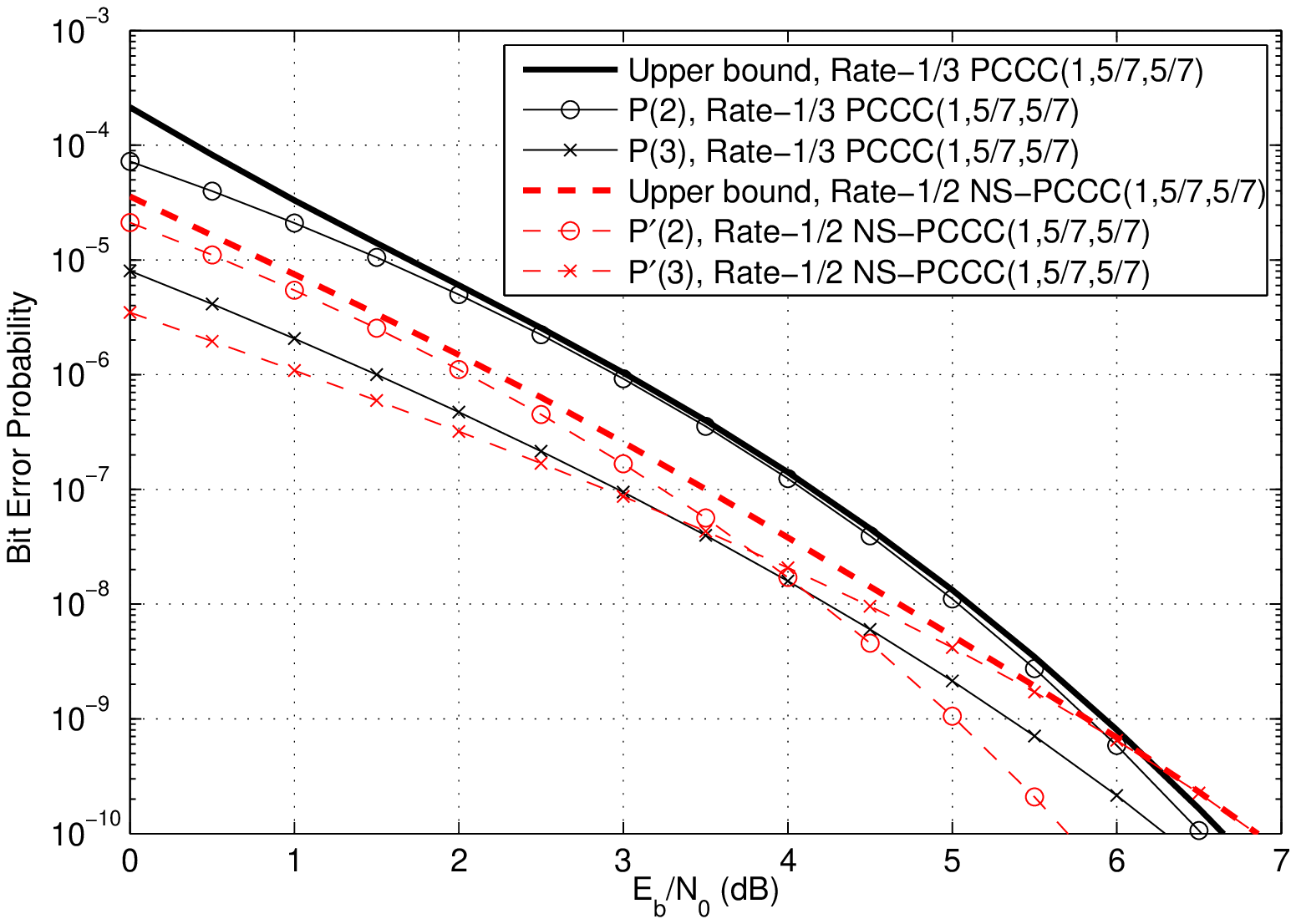}
            \caption{Upper bounds and contributions to the BEP of all error events with information weight of 2 and 3, for the rate-1/2 NS-PCCC(1,5/7,5/7) and its parent rate-1/3 PCCC. The interleaver size is 1,000.}
            \label{Fig_BER_PCCCr5f7_L1000}
        \end{minipage}
        \\
        \begin{minipage}[b]{0.96\linewidth}
            \centering
            \includegraphics[width=1\linewidth]{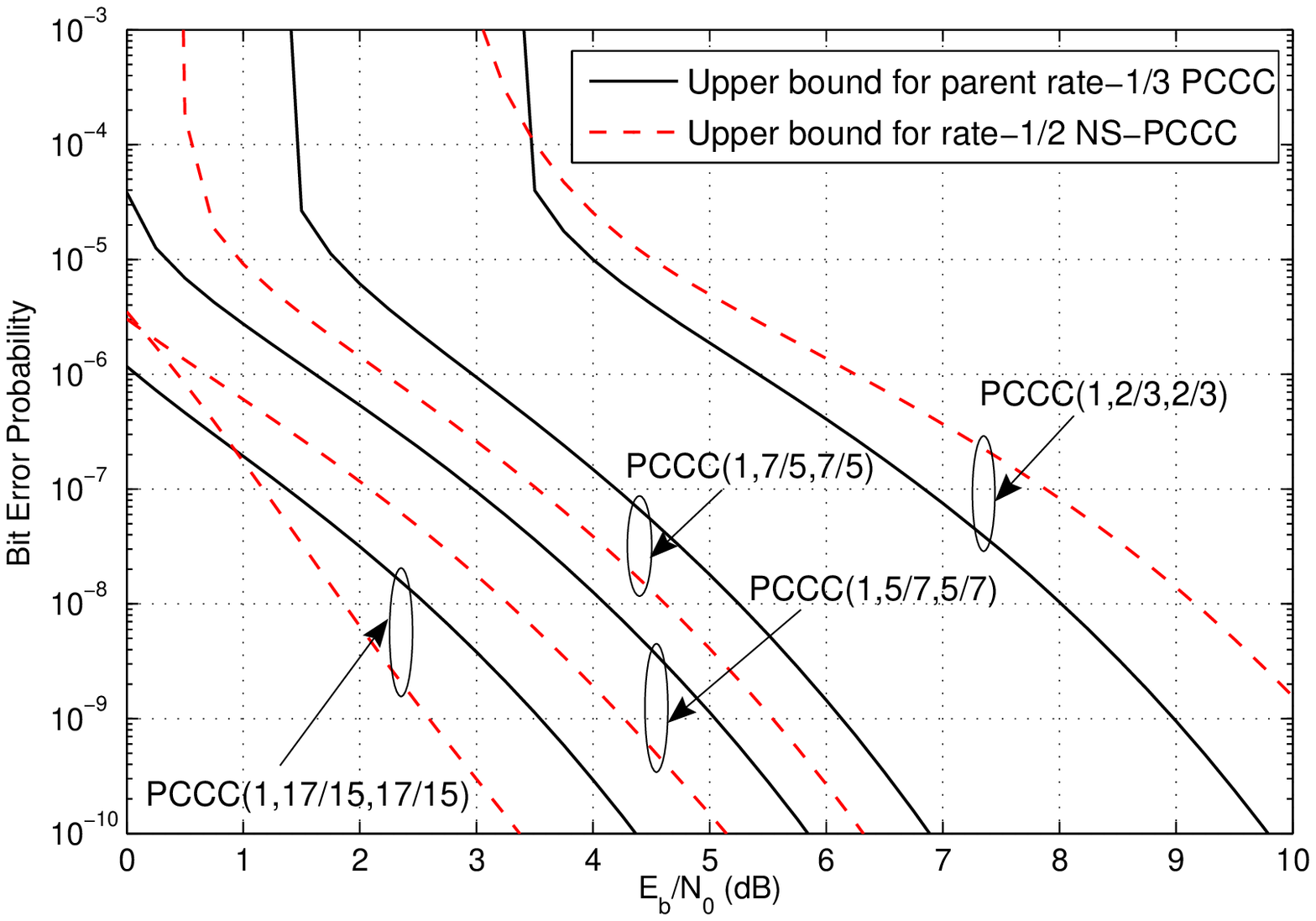}
            \caption{Comparisson of upper bounds for various rate-1/2 NS-PCCCs and their parent rate-1/3 PCCCs. The interleaver size is 10,000.}
            \label{Fig_BER_L10000}
        \end{minipage}
    \end{tabular}
\end{figure}

The contributions of the error events with weight 2 and 3, for the
previous case, are examined in Fig.\ref{Fig_BER_PCCCr5f7_L1000}. We
observe that up to a certain value of $E_{b}/N_{0}$, the rate-1/2
PCCC(1,5/7,5/7) exhibits a lower upper bound than its parent code.
Note that for this range of $E_{b}/N_{0}$ values, the inequality
$P^{\prime}(2)<P(2)$ is satisfied. Although $P(3)$ and
$P^{\prime}(3)$ do not significantly affect the bounds at low
$E_{b}/N_{0}$ values, they play an important role at higher
$E_{b}/N_{0}$ values. However, for a larger interleaver size (e.g.,
$N=10,000)$, $P^{\prime}(2)$ is dominant, as explained previously.
More specifically, $P^{\prime}(2)$ determines the upper bound of the
rate-1/2 NS-PCCC(1,5/7,5/7), which is always lower than the upper
bound of its parent rate-1/3 PCCC for the range of $E_{b}/N_{0}$
values investigated, as we observe in Fig.\ref{Fig_BER_L10000}.

\begin{figure*}[t]
  \centerline{
    \subfigure[Rate-1/3 PCCC(1,5/7,5/7)]{\label{fig:EXIT-Sys}\includegraphics[width=0.33\linewidth]{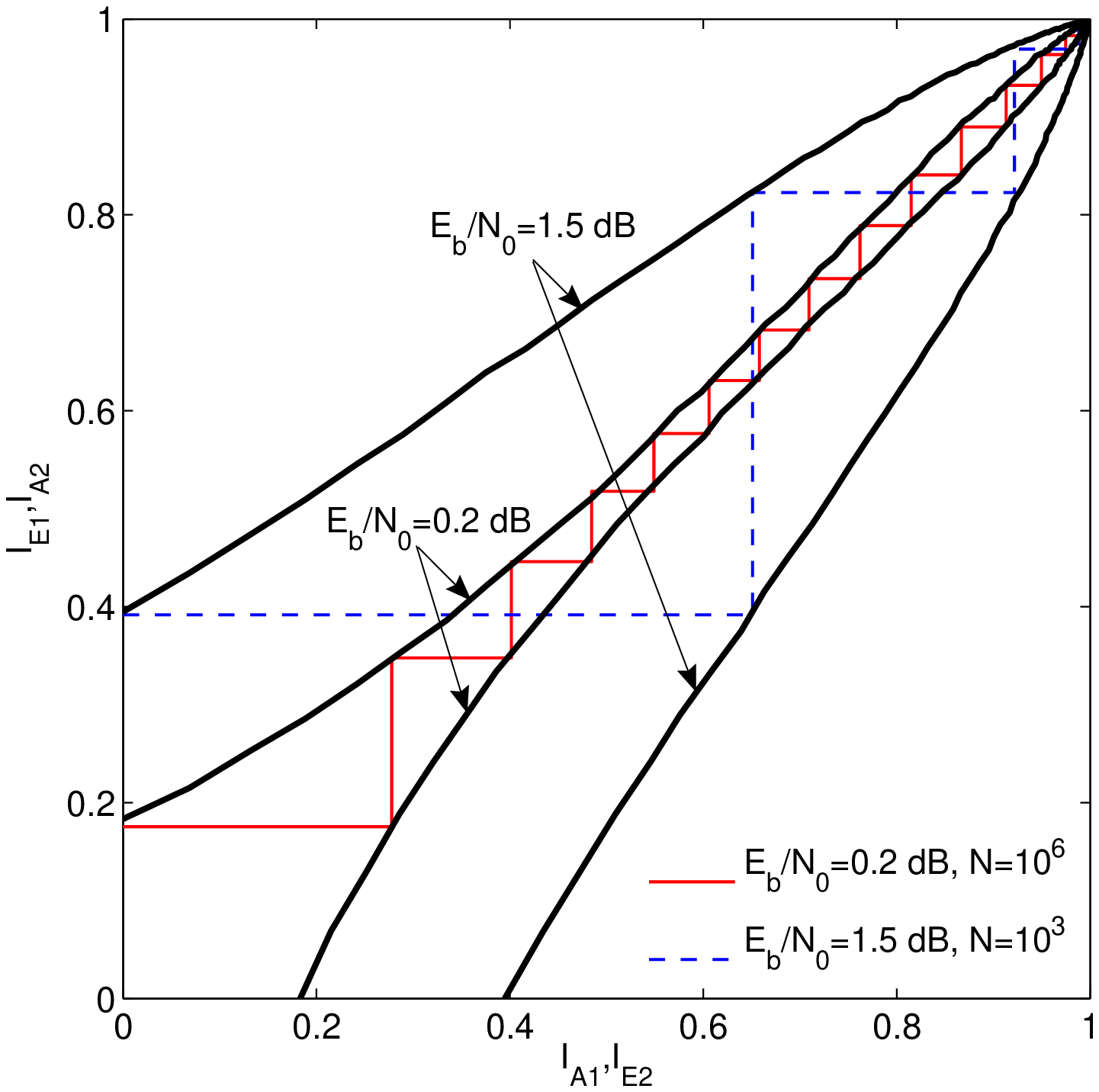}}
    \hfil
    \subfigure[Rate-1/2 NS-PCCC(1,5/7,5/7)]{\label{fig:EXIT-NonSys}\includegraphics[width=0.33\linewidth]{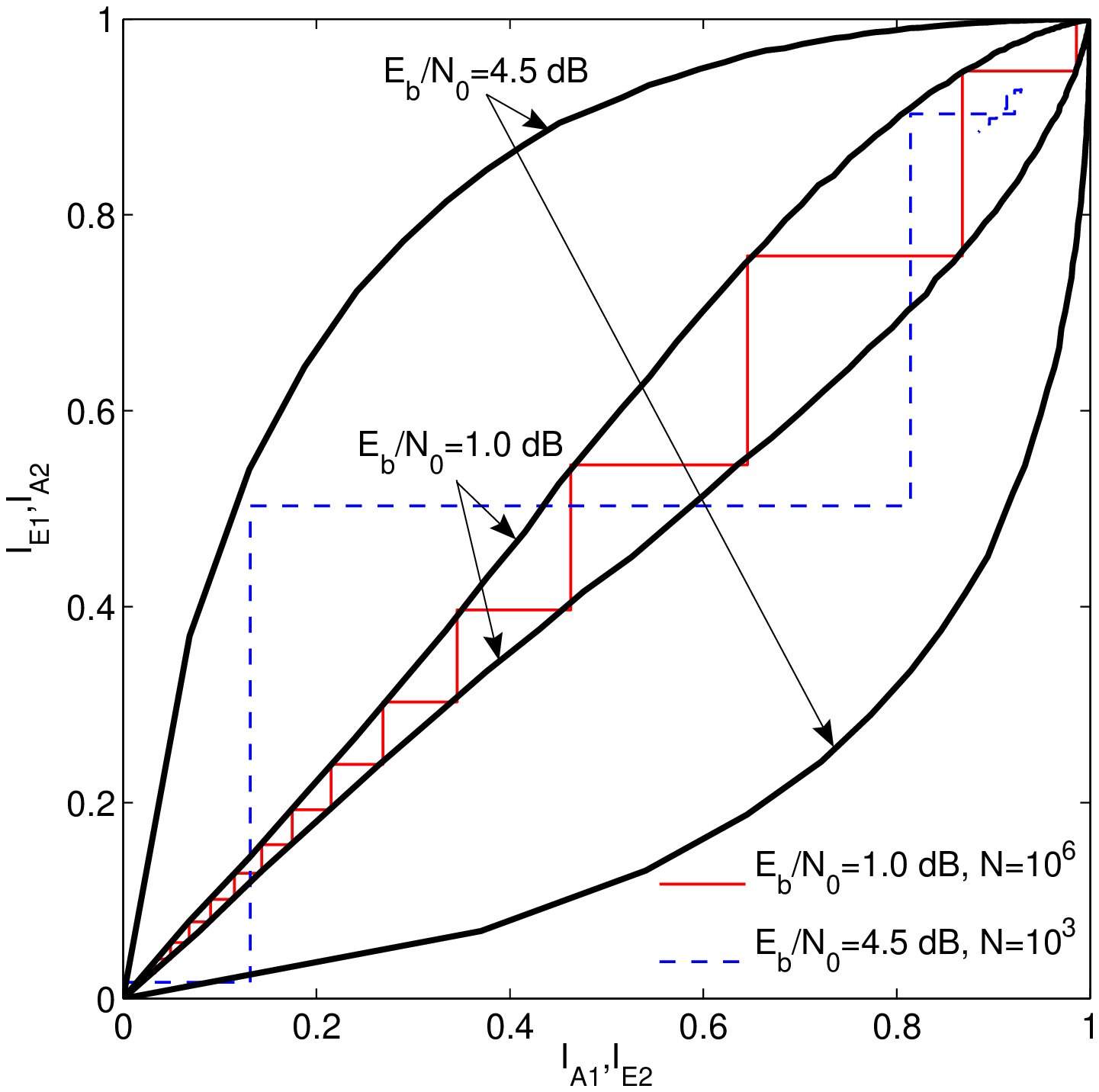}}
    \hfil
    \subfigure[Rate-1/2 PS-PCCC(1,5/7,5/7)]{\label{fig:EXIT-PartSys}\includegraphics[width=0.33\linewidth]{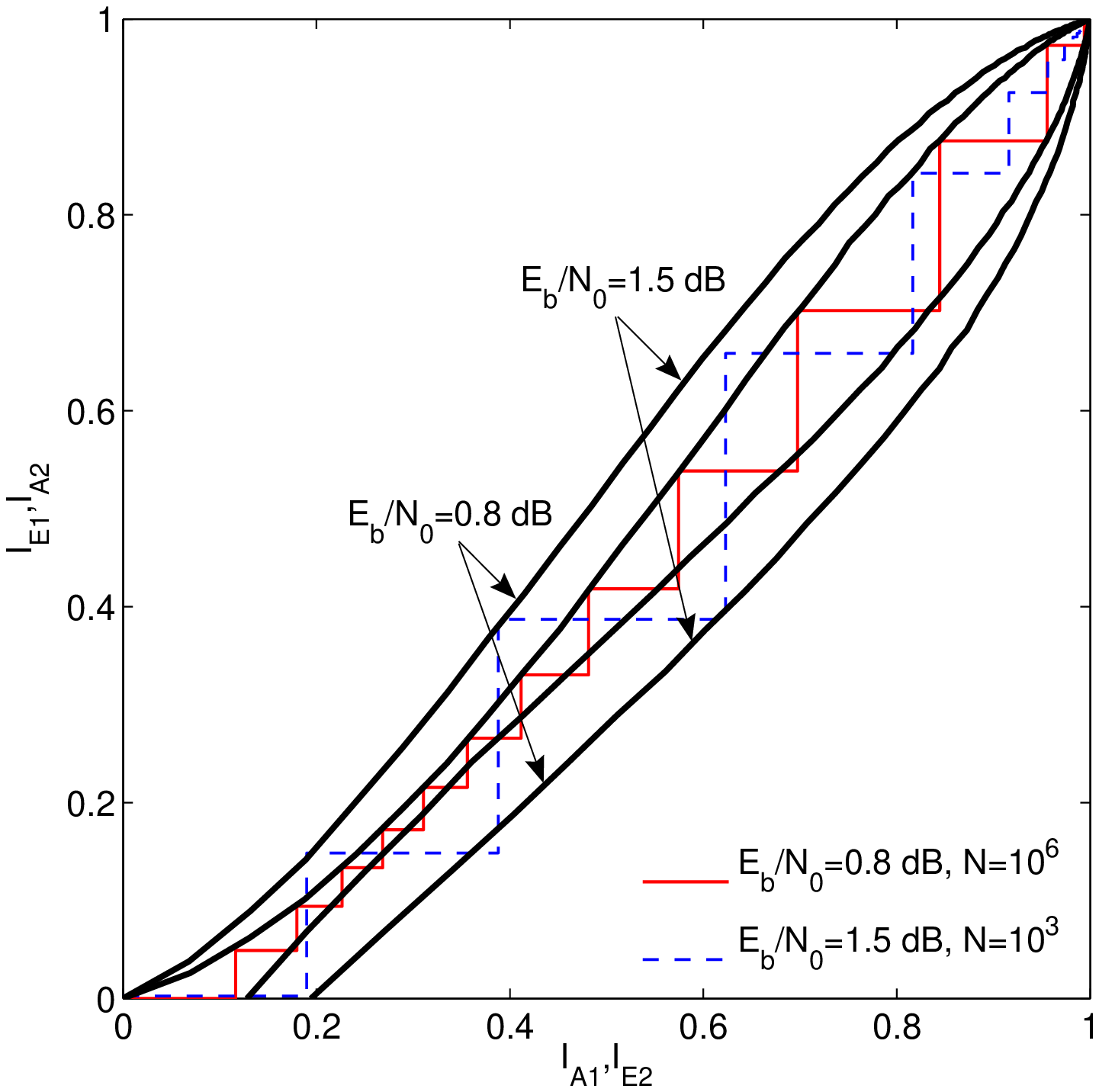}}
  }
  \caption{Extrinsic information transfer characteristics of iterative decoding for various turbo codes using an interleaver size of $10^6$ bits. Two decoding averaged trajectories for each case are also plotted; one for an interleaver size of $10^3$ bits and the other for an interleaver size of $10^6$ bits.}
  \label{EXIT-Charts-1}
\end{figure*}

In Fig.\ref{Fig_BER_L10000}, we also observe that in all cases
except for PCCC(1,2/3,2/3), the rate-1/2 NS-PCCCs achieve better
bounds than their parent PCCCs. The reason is that the free
effective distance of the parent rate-1/3 PCCC(1,2/3,2/3) is
$d_{\text{free.eff}}=4$, thus condition (\ref{Find_Condition_6}) is
not met. For all other turbo codes investigated in
Fig.\ref{Fig_BER_L10000}, $d_{\text{free.eff}}>6$ holds true.


\section{Convergence Considerations}
\label{Convergence}

The upper bound on the BEP for ML soft decoding provides an accurate
estimate of the suboptimal iterative decoder performance at high
$E_{b}/N_{0}$ values, for an increasing number of iterations
\cite{Benedetto96a}. Since the performance of rate-1/3 PCCCs
gradually converges to the ML bound, this bound can be used to
predict the BEP error floor region of the corresponding code.
However, when puncturing occurs, we need to explore whether the
performance of the iterative decoder eventually converges to the ML
bound. For this reason, we use extrinsic information transfer (EXIT)
chart analysis \cite{TenBrink01}, which can accurately predict the
convergence behavior of the iterative decoder for very large
interleaver sizes (e.g., $N\!=\!10^6$ bits).


An iterative decoder consists of two soft-input/soft-output
decoders. Each decoder uses the received systematic and parity bits
as well as a-priori knowledge from the previous decoder to produce
extrinsic information on the systematic bits. Ten Brink described
the decoding algorithm process using EXIT chart analysis
\cite{TenBrink01}. To this end, the information content of the
a-priori knowledge is measured using the mutual information $I_{A}$
between the information bits at the transmitter and the a-priori
input to the constituent decoder. Mutual information $I_{E}$ is also
used to quantify the extrinsic output. The extrinsic information
transfer characteristics are then defined as a function of $I_{A}$
and $E_{b}/N_{0}$, i.e.,
$I_{E}\!=\!T\left(I_{A},E_{b}/N_{0}\right)$. By plotting the mutual
information transfer characteristics of both constituent decoders in
a single EXIT chart, evolution of the iterative decoding process can
be visualised.

During the first iteration, the first decoder does not have any
a-priori knowledge, thus $I_{A_{1},1}\!=\!0$, while the second
decoder uses the extrinsic output
$I_{E_{1},1}\!=\!T\left(0,E_{b}/N_{0}\right)$ of the first decoder
as a-priori knowledge, i.e., $I_{A_{2},1}\!=\!I_{E_{1},1}$. The
extrinsic output of the second decoder,
$I_{E_{2},1}\!=\!T\left(I_{A_{2},1},E_{b}/N_{0}\right)$, is
forwarded to the first decoder to become a-priori knowledge during
the next iteration, i.e., $I_{A_{1},2}\!=\!I_{E_{2},1}$, and so on.
Note that convergence to the $(I_{A},I_{E})\!=\!(1,1)$ point, i.e.,
towards low BEPs, occurs if the transfer characteristics do not
cross.

As an example, we consider the rate-1/3 systematic PCCC(1,5/7,5/7)
to be the parent code. Fig.\ref{fig:EXIT-Sys} shows the transfer
characteristics of the constituent decoders for the parent PCCC
using an interleaver of size $N\!\!=\!\!10^6$. We see that for
$E_{b}/N_{0}\!=\!0.2$ dB, the decoder characteristics do not
intersect and the averaged decoding trajectory \cite{TenBrink01}
manages to go through a narrow tunnel. Fig.\ref{fig:EXIT-NonSys}
depicts the decoder characteristics for the rate-1/2 NS-PCCC
employing an interleaver of the same size. For $E_{b}/N_{0}\!=\!1.0$
dB, the averaged trajectory just manages to pass through a narrow
opening, which appears close to the starting point $(0,0)$.
Therefore, for long interleavers, the performance of the suboptimal
iterative decoder for the rate-1/2 NS-PCCC converges towards the
error floor region, defined by the upper bound, and eventually
outperforms its rate-1/3 parent PCCC. However, convergence begins at
a higher $E_{b}/N_{0}$ value and a larger number of iterations is
required.

In Fig.\ref{fig:EXIT-Sys} and Fig.\ref{fig:EXIT-NonSys} the averaged
trajectories for the more practical interleaver size of 1,000 bits
are also depicted. For $E_{b}/N_{0}\!=\!1.5$ dB the trajectory for
rate-1/3 PCCC quickly converges towards low BEPs. However, the
trajectory for rate-1/2 NS-PCCC dies away after 2 iterations, even
for an $E_{b}/N_{0}$ value of 4.5 dB, due to the increasing
correlations of extrinsic information. We attribute this problem to
the absence of received systematic bits, which causes erroneous
decisions. As a result, error propagation prohibits the iterative
decoder from converging. Thus, for small and more practical
interleaver sizes, the performance of rate-1/2 NS-PCCC does not
gradually approach the upper bound for ML decoding.

To alleviate this problem, the turbo encoder could send some
systematic bits, while keeping the rate equal to 1/2 by puncturing
some parity bits. In \cite{Chatzigeorgiou06b} we have presented a
technique for deriving good punctured codes and we have identified a
rate-1/2 PS-PCCC that achieves the second best performance bound for
ML decoding, after the rate-1/2 NS-PCCC. Fig.\ref{fig:EXIT-PartSys}
shows the transfer characteristics of the constituent decoders as
well as the averaged trajectory for $N\!\!=\!\!10^6$. A comparison
with Fig.\ref{fig:EXIT-NonSys} for the NS-PCCC case, reveals that a
lower $E_{b}/N_{0}$ is required and less iterations are needed, in
order for the rate-1/2 PS-PCCC to converge. Furthermore, the
trajectory for an interleaver size of 1,000 bits reaches the top
corner of the EXIT chart for the same $E_{b}/N_{0}$ as the parent
rate-1/3 PCCC, guaranteeing that the iterative decoder will converge
towards low BEP values.


\section{Simulation Results}
\label{results}

The process for selecting rate-1/2 PS-PCCCs that lead to superior
performance than their parent rate-1/3 PCCCs can be summarized in
two steps. We first implement the technique we have proposed in
\cite{Chatzigeorgiou06b} to derive good punctured PCCCs that exhibit
low error floors. We then use EXIT charts and averaged trajectories
to identify the punctured PCCC whose performance converges towards
the error floor region, when iterative decoding is used.

\begin{figure}[t]
    \centering
    \includegraphics[width=1\linewidth]{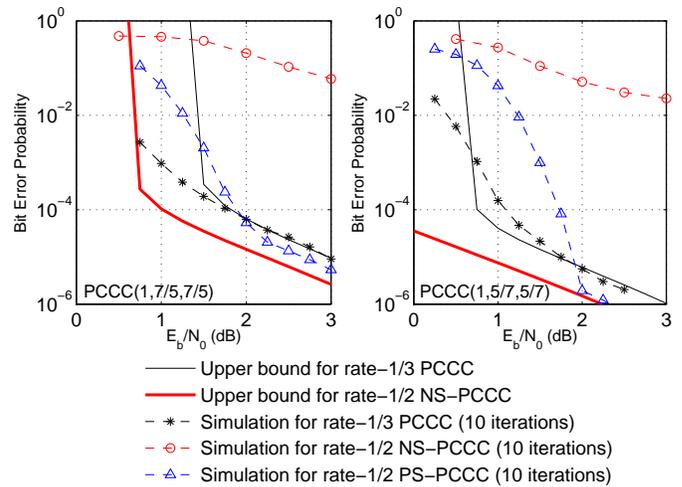}
    \caption{Comparison between the upper bounds and simulation results for PCCCs using an interleaver size of 1,000 bits.}
    \label{Fig_Simulations}
\end{figure}

We consider PCCC(1,5/7,5/7) and PCCC(1,7/5,7/5) to demonstrate the
effectiveness of this process. The iterative decoder applies the
BCJR algorithm \cite{Bahl74} and performance is plotted after 10
iterations. A random interleaver of size 1,000 bits is used. In
Fig.\ref{Fig_Simulations} we see that the performance of both parent
rate-1/3 PCCCs coincides with the corresponding upper bounds for ML
decoding, at high $E_{b}/N_{0}$ values. As expected, the performance
of iterative decoding for rate-1/2 NS-PCCCs does not converge
towards low BEPs, thus they do not outperform their parent codes,
although they exhibit a lower upper bound. Nevertheless, rate-1/2
PS-PCCCs that achieve a lower error floor than their parent rate-1/3
PCCCs, can be found based on \cite{Chatzigeorgiou06b}. Note that, in
both cases, the encoder of the selected rate-1/2 PS-PCCC transmits 7
parity bits and only 1 systematic bit for every 4 input information
bits.

\section{Conclusion}
\label{conclusion}

In this paper we have demonstrated that, if a certain condition is
met over the AWGN channel, puncturing of the systematic output of a
rate-1/3 turbo code using a random interleaver leads to a rate-1/2
non-systematic turbo code that achieves a better performance than
its parent code, when ML decoding is used. In the case of iterative
decoding, the absence of systematic bits makes convergence towards
low bit-error rates difficult for the rate-1/2 non-systematic turbo
decoder. Nevertheless, we can assist convergence by transmitting
some systematic bits, while keeping the rate 1/2 by puncturing more
parity bits. Thus, we can combine the techniques described in
\cite{Chatzigeorgiou06b} and \cite{TenBrink01} to identify good
puncturing patterns, improve bandwidth efficiency by reducing the
rate of a PCCC from 1/3 to 1/2 and, at the same time, achieve a
lower error floor.

\bibliographystyle{IEEEtran}
\bibliography{IEEEabrv,TurboRefShort}

\begin{thebibliography}{10}
\providecommand{\url}[1]{#1}
\csname url@rmstyle\endcsname
\providecommand{\newblock}{\relax}
\providecommand{\bibinfo}[2]{#2}
\providecommand\BIBentrySTDinterwordspacing{\spaceskip=0pt\relax}
\providecommand\BIBentryALTinterwordstretchfactor{4}
\providecommand\BIBentryALTinterwordspacing{\spaceskip=\fontdimen2\font plus
\BIBentryALTinterwordstretchfactor\fontdimen3\font minus
  \fontdimen4\font\relax}
\providecommand\BIBforeignlanguage[2]{{%
\expandafter\ifx\csname l@#1\endcsname\relax
\typeout{** WARNING: IEEEtran.bst: No hyphenation pattern has been}%
\typeout{** loaded for the language `#1'. Using the pattern for}%
\typeout{** the default language instead.}%
\else
\language=\csname l@#1\endcsname
\fi
#2}}
\renewcommand\BIBentryALTinterwordstretchfactor{4}

\bibitem{Hagenauer88}
J.~Hagenauer, ``Rate compatible punctured convolutional codes and their
  applications,'' \emph{{IEEE} Trans. Commun.}, vol.~36, pp. 389--400, Apr.
  1988.

\bibitem{Haccoun89}
D.~Haccoun and G.~B{\'{e}}gin, ``High-rate punctured convolutional codes for
  {V}iterbi and sequential decoding,'' \emph{{IEEE} Trans. Commun.}, vol.~37,
  pp. 1113--1125, Nov. 1989.

\bibitem{Acikel99}
{\"{O}}.~A{\c{c}}ikel and W.~E. Ryan, ``Punctured turbo-codes for {BPSK/QPSK}
  channels,'' \emph{{IEEE} Trans. Commun.}, vol.~47, pp. 1315--1323, Sept.
  1999.

\bibitem{Babich02}
F.~Babich, G.~Montorsi, and F.~Vatta, ``Design of rate-compatible punctured
  turbo ({RCPT}) codes,'' in \emph{Proc. Int. Conf. Comm. ({ICC}'02)}, New
  York, USA, Apr. 2002, pp. 1701--1705.

\bibitem{Kousa02}
M.~A. Kousa and A.~H. Mugaibel, ``Puncturing effects on turbo codes,''
  \emph{Proc. {IEE} Comm.}, vol. 149, pp. 132--138, June 2002.

\bibitem{FanMo99}
M.~Fan, S.~C. Kwatra, and K.~Junghwan, ``Analysis of puncturing pattern for
  high rate turbo codes,'' in \emph{Proc. Military Comm. Conf. ({MILCOM}'99)},
  New Jersey, USA, Oct. 1999, pp. 547--500.

\bibitem{Land00}
I.~Land and P.~Hoeher, ``Partially systematic rate 1/2 turbo codes,'' in
  \emph{Proc. Int. Symp. Turbo Codes}, Brest, France, Sept. 2000, pp. 287--290.

\bibitem{Blazek02}
Z.~Blazek, V.~K. Bhargava, and T.~A. Gulliver, ``Some results on partially
  systematic turbo codes,'' in \emph{Proc. Vehicular Tech. Conf.
  ({VTC}-Fall'02)}, Vancouver, Canada, Sept. 2002, pp. 981--984.

\bibitem{Chatzigeorgiou06b}
I.~Chatzigeorgiou, M.~R.~D. Rodrigues, I.~J. Wassell, and R.~Carrasco, ``A
  novel technique for the evaluation of the transfer function of punctured
  turbo codes,'' in \emph{Proc. Intl. Conf. Comm. ({ICC}'06)}, Istanbul,
  Turkey, July 2006.

\bibitem{Berrou96}
C.~Berrou and A.~Glavieux, ``Near optimum error correcting coding and decoding:
  Turbo codes,'' \emph{{IEEE} Trans. Commun.}, vol.~44, pp. 1261--1271, Oct.
  1996.

\bibitem{Divsalar95}
D.~Divsalar, S.~Dolinar, R.~J. McEliece, and F.~Pollara, ``Transfer function
  bounds on the performance of turbo codes,'' JPL, Cal. Tech., TDA Progr. Rep.
  42-121, Aug. 1995.

\bibitem{Benedetto96a}
S.~Benedetto and G.~Montorsi, ``Unveiling turbo codes: Some results on parallel
  concatenated coding schemes,'' \emph{{IEEE} Trans. Inform. Theory}, vol.~42,
  pp. 409--429, Mar. 1996.

\bibitem{Hall98}
E.~K. Hall and S.~G. Wilson, ``Design and analysis of turbo codes on rayleigh
  fading channels,'' \emph{{IEEE} J. Select. Areas Commun.}, vol.~16, pp.
  160--174, Feb. 1998.

\bibitem{Benedetto96b}
S.~Benedetto and G.~Montorsi, ``Design of parallel concatenated convolutional
  codes,'' \emph{{IEEE} Trans. Commun.}, vol.~44, pp. 591--600, May 1996.

\bibitem{TenBrink01}
S.~ten Brink, ``Convergence behavior of iteratively decoded parallel
  concatenated codes,'' \emph{{IEEE} Trans. Commun.}, vol.~49, pp. 1727--1737,
  Oct. 2001.

\bibitem{Bahl74}
L.~R. Bahl, J.~Cocke, F.~Jelinek, and J.~Raviv, ``Optimal decoding of linear
  codes for minimising symbol error rate,'' \emph{{IEEE} Trans. Inform.
  Theory}, vol. IT-20, pp. 284--287, Mar. 1974.

\end{thebibliography}

\end{document}